 \documentclass[final,3p,times,twocolumn]{elsarticle}

\usepackage{amssymb}

\usepackage[latin1]{inputenc}
\usepackage[T1]{fontenc}
\usepackage{xcolor}
\usepackage[english]{babel}
\usepackage{amsmath} 
\usepackage{natbib}
\usepackage{graphics} 
\usepackage{subcaption}

\usepackage{url} 

\journal{Acta Astronautica}

\begin{document}

\begin{frontmatter}



\title{The First Interstellar Astronauts Will Not Be Human}


\author[inst1,inst2,cor1]{Stephen Lantin}
\author[inst3,inst4]{Sophie Mendell}
\author[inst3]{Ghassan Akkad}
\author[inst5]{Alexander N. Cohen}
\author[inst5]{Xander Apicella}
\author[inst5]{Emma McCoy}
\author[inst6]{Eliana Beltran-Pardo}
\author[inst7]{Michael Waltemathe}
\author[inst8,inst9]{Prasanna Srinivasan}
\author[inst3]{Pradeep M. Joshi}
\author[inst3]{Joel H. Rothman}
\author[inst5]{Philip Lubin}

\cortext[cor1]{Corresponding Author, slantin@ufl.edu}

\affiliation[inst1]{organization={Department of Agricultural and Biological Engineering, University of Florida},
            city={Gainesville},
            postcode={32611}, 
            state={FL},
            country={USA}}
\affiliation[inst2]{organization={Department of Chemical Engineering, University of California - Santa Barbara},
            city={Santa Barbara},
            postcode={93106}, 
            state={CA},
            country={USA}}
\affiliation[inst3]{organization={Department of Molecular, Cellular, and Developmental Biology, University of California - Santa Barbara},
            city={Santa Barbara},
            postcode={93106}, 
            state={CA},
            country={USA}}
\affiliation[inst4]{organization={College of Creative Studies, University of California - Santa Barbara},
            city={Santa Barbara},
            postcode={93106}, 
            state={CA},
            country={USA}}
\affiliation[inst5]{organization={Department of Physics, University of California - Santa Barbara},
            city={Santa Barbara},
            postcode={93106}, 
            state={CA},
            country={USA}}
\affiliation[inst6]{organization={UCLA Health Center},
            city={757 Westwood Plaza, Los Angeles},
            postcode={90095}, 
            state={CA},
            country={USA}}
\affiliation[inst7]{Ruhr-Universitat Bochum, Universitatsstrasse 150, 44801 Bochum, Germany,
            }
\affiliation[inst8]{organization={Department of Electrical and Computer Engineering, University of California - Santa Barbara},
            city={Santa Barbara},
            postcode={93106}, 
            state={CA},
            country={USA}}
\affiliation[inst9]{organization={Center for BioEngineering, University of California - Santa Barbara},
            city={Santa Barbara},
            postcode={93106}, 
            state={CA},
            country={USA}}

\begin{abstract}
Our ability to explore the cosmos by direct contact has been limited to a small number of lunar and interplanetary missions. However, the NASA Starlight program points a path forward to send small, relativistic spacecraft far outside our solar system via standoff directed-energy propulsion. These miniaturized spacecraft are capable of robotic exploration but can also transport seeds and organisms, marking a profound change in our ability to both characterize and expand the reach of known life. Here we explore the biological and technological challenges of interstellar space biology, focusing on radiation-tolerant microorganisms capable of cryptobiosis. Additionally, we discuss planetary protection concerns and other ethical considerations of sending life to the stars.
\end{abstract}



\begin{keyword}
NASA Starlight \sep cryptobiosis \sep directed energy propulsion \sep interstellar propulsion \sep space biology \sep planetary protection
\PACS 87.53.-j \sep 42.62.-b \sep 07.87.+v
\MSC 92C99 \newline
\\
This is a pre-print of an \textit{Acta Astronautica} article, which can be found here: doi.org/10.1016/j.actaastro.2021.10.009
\end{keyword}

\end{frontmatter}



\section{{Propagating} Life over Interstellar Distances}

\subsection{Introduction}

Since the inception of the space programs, thousands of spacecraft have flown around Earth and into the solar system.
However, spacecraft in interstellar space, despite the prospect of new discovery and the promises of science fiction, are
few and far between---less than five have ventured beyond the heliopause. This discrepancy is explained in part by
distance; it takes decades for spacecraft to travel the 18 billion kilometers ($\sim$120 AU) to interstellar space
with traditional chemical propulsion. Project Starlight (DEEP-IN), a NASA Innovative Advanced Concepts (NIAC) program
promises to both radically increase maximum spacecraft speed and redefine the economics of space
propulsion~\citep{Lubin2016}. This requires the multi-decade development of a system that employs a large-aperture
standoff directed-energy laser-phased array, henceforth referred to as a ``directed energy (DE) system''. By illuminating
a reflector (laser sail) mounted to a spacecraft, it is possible to achieve velocities that are significant fractions of
the speed of light. The key is to leave the ``propulsion at home'' and use direct photon momentum exchange to achieve
relativistic speeds. What took the \textit{Voyager} and \textit{Pioneer} space probes decades to travel to interstellar
space will take the next generation of spacecraft days---and without the need for on-board propellant.

The reason we have not had any serious proposals for relativistic flight to achieve interstellar capability is because no realistic technology has existed until recently. All chemical and nuclear propulsion strategies lack the energy per mass to achieve relativistic flight. Only antimatter and photon propulsion are consistent with current physics and engineering, and antimatter lacks engineering options for realization. This leaves only photon propulsion as a realistic option. Driven by a wide variety of needs, photonics, like electronics, has grown exponentially in capability and dropped exponentially in cost with a doubling time in capability and cost reduction of about two years. Project Starlight uses this exponential growth in photonics as the basis for the program.

With relativistic flight on the horizon, the question remains: Why should we develop the technology to send spacecraft
into interstellar space? The answer lies in two general areas. One is the human drive to understand and explore, which is
partly logical and not easily expressed in logic. The other reason is vastly more practical: the same technology we use to
enable relativistic flight also enables a transformative range of possibilities for space exploration. This includes rapid
high-mass transport in space and human-capable missions in the solar system and beyond. Other benefits, such as planetary
defense options and extremely large unique optical phased array telescopes, are also foreseen. From the perspective of
exoplanet exploration, sending small telescopes close to exoplanets allows vastly higher spatial resolution than we can
achieve with conceivable Earth or space-based telescopes~\citep{Lubin2016}.
The combination of spacecraft swarming -- launching large numbers of small probes -- with earth and orbital assets is a synergistic approach to exploration.

\subsection{What is Interesting to Send Out?}

What is interesting to send out to interstellar space? There is precedent for deliberately sending out information to the
stars, an action known as Active SETI (Search for Extraterrestrial Intelligence) or messaging to extraterrestrial
intelligence (METI). In 1972 and 1973, \textit{Pioneer 10} and \textit{Pioneer 11}, respectively, were launched into space
with plaques depicting the human form, the location of the solar system and Earth in relation to pulsars, and more.
Similarly, the Arecibo message, a radio transmission sent to globular cluster M13 in 1974, reproduced the human form and
Earth's location. Three years later, the \textit{Voyager} spacecrafts carried golden records into space, which included
sounds and images from Earth. In 2017 and 2018, sets of ten second audio transmissions were sent to Luyten's star in the
art project S\'onar Calling GJ273b. The contents of Active SETI messages -- and the notion that they should be sent out at
all -- is highly debated; the inquisitive reader will find more detail in an overview by Musso~\citep{Musso2012}. 
Like the artifacts of human societies separated from us by time, the contents of a laser-sail spacecraft may provide the only insight into humankind. Not only will it be deeply reflective of our inherent values as a species, but the information may direct extraterrestrial life's first impression of humanity.

 While previous interstellar missions have sent messages, the options for what can be sent via a swarm of laser-sail spacecraft are wider in scope. With the cost of fabricating each spacecraft decreasing over time and aside from an initial upfront investment to build the array, it will be relatively inexpensive to send out multiple spacecraft.  \textbf{Together with a thousand-fold decrease in the time it takes to reach extrasolar space, life -- particularly living microorganisms -- can be studied in the interstellar space environment for the first time.} Launching ``biosentinel'' experiments into empty interstellar space provide research possibilities that will be able to more fully characterize biological systems and pave the way for human travel to the stars.

In this work, we adopt the definition of life provided by NASA's Astrobiology Institute: that which is ``a self-sustaining
chemical system capable of Darwinian evolution''~\citep{Benner2010DefiningLife.}.
Understandably, this is limited by our terrestrial context and is subject to change in the future given a nuanced or
drastically different version of life is found outside of Earth. For more complex life, including \textit{Homo sapiens},
to safely embark on interplanetary and interstellar missions, we must know more about life's ability to endure the
deep-space environment. However, we cannot obtain relevant data from ground-based studies and in low-Earth orbit (LEO).
While instruments for simulating microgravity exist, such as ground-based clinostats and rotating wall vessel bioreactors,
they cannot fully reproduce the lack of structural deformation, displacement of intracellular components, or reduced mass
transfer associated with the proposed spaceflight environment~\citep{Horneck2010}.
These experiments also lack the coupled effects of microgravity and space radiation, which may elicit different biological
responses than exposure to only one stressor~\citep{Moreno-Villanueva2017}. LEO studies do not provide a suitable
experimental environment for understanding biological responses to interplanetary/interstellar radiation either; the
radiation dose rate is substantially higher in interplanetary space than inside LEO (i.e.,~in the International Space
Station (ISS)) with comparable shielding, and the quality factor ($Q$) is substantially larger outside of LEO with
light shielding~\citep{Straume2017}. Thus, only biosentinel experiments can provide the necessary data.

\section{Engineering Vision and Challenges}

Starlight's directed-energy system is both modular and scalable, allowing a logical development and build progression from very low-mass wafer-scale spacecraft to much larger and more capable spacecraft. Even small wafer-scale spacecraft attached to a laser sail (Figure~\ref{fig:DEEP-laser-sail}) present a range of options that we explore here. As Starlight is not limited to small spacecraft, the long-term future opens this technology to a variety of mission options. The initial small spacecraft become scouting or sentinel missions much like our solar system exploration programs. The space environment and the interstellar distances present engineering challenges which demand reliable solutions for the success of interstellar biosentinel experiments. These challenges include the need for propulsion and robust communication, which influence payload selection and experimental design.

\subsection{Propulsion}

\begin{figure*}
\begin{subfigure}{\textwidth}
    \centering
        \centering
        \includegraphics[width=0.65\textwidth]{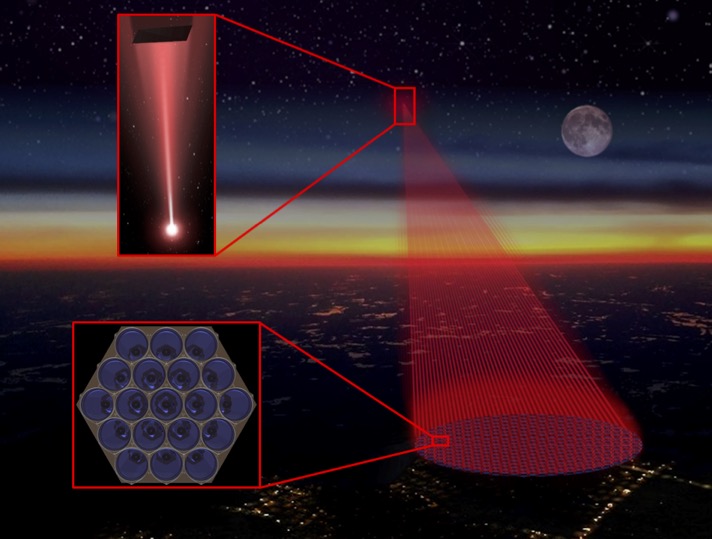} 
        \caption{}
        \label{fig:DEEP-laser-sail}
\end{subfigure}
    \hfill
\begin{subfigure}{\textwidth}
        \centering
        \includegraphics[width=0.65\textwidth]{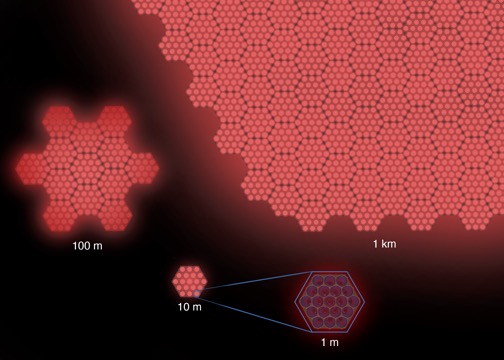} 
        \caption{}
        \label{fig:DE-STAR}
\end{subfigure}
    \caption{\bf{Directed Energy Propulsion of a Light Sail.} \bf{(a) }\normalfont{A light sail and payload propelled into interstellar space by directed energy laser propulsion. Emitted photons from a standoff laser array on the surface of the Earth (space-based laser arrays are also possible) impart momentum on the sail by reflection so as to accelerate the spacecraft up to relativistic speeds.} \textit{Artist's rendition.} \bf{(b) }\normalfont{The laser array is composed of many small, modular sub-elements which can be articulated, switched off, and added so as to enable a large mission space. As the capability of directed energy propulsion grows, relativistic flight will become possible.}}
\end{figure*}

The NASA Starlight program has developed a standoff laser-phased array that is scalable to sizes needed to provide relativistic flight (Figure~\ref{fig:DEEP-laser-sail}). Since the primary directed-energy system is modular and scalable from handheld to kilometer scale, the program allows a ``roadmap to interstellar flight'' with milestones and applications at each step. A system capable of achieving relativistic flight for small spacecraft at 0.25\textit{c}, for example, would require a directed energy power level on the order of 100 GW for a few minutes per launch for wafer scale spacecraft, with a DE array size scale of order 1--10~km. The power scale is large (about 1/10 of the US electrical grid) but it is only used for a few minutes per launch for low-mass missions. The required level of energy needed for a launch is about 0.01 percent of the US energy used per day. The ``roadmap'' proposed starts with Earth-based systems, such as those we propose in the NASA Starlight program, and will evolve to orbital and lunar-based systems. This is a long-term transformational program, not a single-use program. Initial variants that are 1--10~m DE arrays are immediately useful for powering and propelling (via DE coupled ion engines) CubeSats, as well as for long-distance power transfer to LEO and geostationary Earth orbit (GEO) assets. They will also be useful against space debris and exploration of the beginning of planetary defense applications. Since the DE array is scalable and each module is identical, mass production and economies of scale will greatly aid the program. In addition to the exponential decrease in cost with time, the cost in large quantities is also lower. This is much like the cost of computation, which has dropped by more than a factor of a million per unit of performance (FLOP) over the past 40~years.

\subsection{Communication}

Data acquisition from interstellar experiments requires a robust downlink (spacecraft-to-Earth) with a baseline laser
communications link. This poses further challenges to successful biosentinel experiment execution, including the severe
limitations on the spacecraft's available electrical power, processing, and transmit systems.  Other challenges include
high propagation loss from diffraction, interfering background radiation, and ground-based reception issues related to
atmospheric turbulence, scattering, and weather. To avoid data volume limitations, careful attention must be paid to: (1)
rendering the communication link as energy efficient as possible, (2) aggressive compression of the data, and (3) use of
reception optics at larger aperture sizes~\citep{Lubin2016,Messerschmitt2018}. Larger distances and
power losses may be compensated by transmitting at a lower data rate and longer times, with a possible mission design to
include a nuclear battery. Such a nuclear battery employing a traditional radioisotope thermoelectric generator (RTG) or thermophotovoltaics  can be designed to fit the ultra-low mass
requirements~\citep{Pustovalov1999,Wilt2007,Hein2017}. Sub gram Pu-238 RTG's we successfuly used in human pacemakers lasting many decades. A data uplink (Earth-to-spacecraft) is also possible and may be
required to configure the transmit wavelength for variations in spacecraft speed. To increase reliability, a swarm of
small wafer-scale spacecraft ($>$1000) can be launched, increasing the redundancy at relatively minimal economic burden with
each additional spacecraft. A highly selective optical bandpass filter will mitigate the radiation from the Earth's
atmosphere (if ground-based), our solar system dust emission and scattering, and nearby stars to improve communication.
Communications issues are further discussed in various papers from within the directed energy
community~\citep{Lubin2016,Messerschmitt2018}.

\section{Physiological Challenges Posed by the Space Environment}

The space environment poses unique challenges for the survival of biological systems and their study. Lifeforms sent into interstellar space will be exposed to microgravity, hypergravity (during launch and propulsion phases), and space radiation. In the absence of proper environmental control, exposure to vacuum and extreme temperatures is also possible. Here, we briefly review relevant ground-based and space-based experiments to aid the development of experimental design heuristics for interstellar space biology. 

\subsection{Microgravity}

In microgravity, convection is driven predominantly by surface tension, resulting in unique patterns of heat and mass
transfer~\citep{Napolitano1984}.
Such environmental conditions give rise to stress responses in several species of microorganisms, including the formation
of biofilms in \textit{Pseudomonas aeruginosa}~\citep{Kim2013}. 
It is well known that biofilms promote the survivability of cells by allowing nutrients to be shared and protected from external hazards such as desiccation and antibiotics. Species that are more suited to biofilm formation (e.g.,~species of bacteria, fungi, and protists) may at present fare better for interstellar travel. 

Another consequence of surface-driven convection is a decrease in fluid mixing, which has been shown to induce local
anoxia~\citep{Briarty2004}.
In bacterial species, convection elicits a shortened period of adjustment to the environment (lag phase) and increased
final cell count when compared with ground-based controls~\citep{Horneck2010}.
However, biological responses associated with decreased fluid mixing are not observed in bacterial species if the growth
medium and strain are conducive to cell motility (i.e.,~by the movement of flagella, etc.)~\citep{Benoit2007}.
In cells of higher organisms, such as plants and animals, a common stress response to microgravity is the downregulation
of genes controlling reactive oxygen species (ROS) levels~\citep{Mahalingam2003}.
Heightened ROS levels in microgravity have been observed in \textit{Brassica napus} (rapeseed) protoplast cells as well as
in \textit{Ceratopteris richardii} (fern) spores, thought to be a result of decreased peroxidase
activity~\citep{Salmi2008}.
Other notable responses include changes in cell membrane structure and function, transcriptome, proteome, cell wall
structure, and Ca$^{2+}$ signaling~\citep{Kordyum2017}.

\subsection{Hypergravity}

While microgravity is characteristic of the deep space environment, spacecraft launch and directed-energy propulsion will
expose biological systems to hypergravity. Microgravity research on Earth requires setups such as drop tubes and parabolic
flights, whereas hypergravity experiments require only a centrifuge, which can simulate gravitational accelerations within
the range of our spacecraft ($10^4$ to $10^6$ g). Tardigrade species are adversely affected by hypergravity but
are more resistant than larger organisms, such as the fruit fly, \textit{D. melanogaster}, while species like \textit{C.
elegans} fare quite well under hyperacceleration~\citep{LeBourg1999,DeSouza2018,Vasanthan2017}.
However, tardigrades can be launched in a cryptobiotic state, with the hope that this will mitigate deleterious
acceleration-dependent effects of hypergravity on their survival rate~\citep{Vasanthan2017}.

\subsection{Space radiation}

We do not envision biological payloads spending extended periods of time in LEO, but we must consider the effects of
radiation during launch. There are three sources of radiation within the low-Earth orbit environment: galactic cosmic
radiation (GCR), solar cosmic radiation (SCR), and trapped radiation as a result of GCR and SCR interactions with Earth's
magnetic field and atmosphere~\citep{Reitz2008}.

Outside of low-Earth orbit, where the vast majority of exposure occurs, there are two sources of significant radiation
concerns. The high-energy GCRs consist of a nucleus of nearly every element on the periodic table as well as electrons,
gamma rays, and neutrinos. The distribution is largely a power law with flux peaking around 1 GeV/nucleon and decreasing
rapidly at higher energies. The primary biological damage, in the form of DNA damage and increased cytotoxicity, comes
from free protons rather than higher-Z nuclei because protons are more numerous, with some studies showing survival rates
of cells falling to 10\% around 5~Gy~\citep{AlanMitteer2015ProtonSpecies}.
The approximate dose from GCR protons is about 6 rad/yr and is extremely difficult to shield against due to their high
energy and penetration; however, the dose is modest enough to not be catastrophic over the multi-decade missions required.
The more serious source of radiation is unique to relativistic missions due to the high-speed collisions with interstellar
medium (ISM). The ISM is dominated by protons and electrons with a density of about 0.25/cc as well as heavier elements at
a much lower density~\citep{Draine2010}.
The ISM particles are largely in thermal equilibrium with a temperature of approximately 8000 K. The particle energies in
the ISM rest frame are very low (${\sim}1~{\rm eV}$) and are not a radiation hazard, but when a relativistic spacecraft impacts
the ISM particles, the radiation hazard on the forward edge becomes extreme with a nearly mono-velocity distribution in
the frame of the spacecraft. This ISM ``impact radiation'' is extremely angle-dependent on the frame of the
spacecraft~\citep{Lubin2020,Cohen2020}.
It is analogous to driving through the rain. For this reason the spacecraft is thin and oriented edge-on like a
``Frisbee'' with the forward edge receiving an extremely large radiation dose (Gigarad/year), while the radiation exposure
on the rearward portion of the spacecraft (where the electronics and biology are located) receive vastly less radiation
that allows a mission to proceed~\citep{Brashears2016}. Further research is needed to confirm if this orientation is
self-stabilizing.

\subsection{Vacuum}

An interstellar mission may have both hermetic enclosures as well as areas open to vacuum. For these apparatus designs, or
in the event of hermetic failure, vacuum compatibility may also help determine experimental viability of species in
interstellar space. Two species of lichen, \textit{Rhizocarpon geographicum} and \textit{Xanthoria elegans}, were found to
withstand the full space environment for two weeks, with no permanent loss of photosynthetic
capabilities~\citep{Sancho2007}. 
The tardigrade species \textit{Richtersius coronifer} and \textit{Milnesium tardigradum}, as well as \textit{Artemia}
brine shrimp cysts, were also found to be vacuum-resistant, albeit for less
time~\citep{Jonsson2008,Hernandorena1999AEnvironment}.
It is important to note that the coupled effect of space radiation with vacuum exposure significantly decreased survivability of the tardigrade species.

\begin{figure*}[h]
\centering
\includegraphics[width=0.75\textwidth]{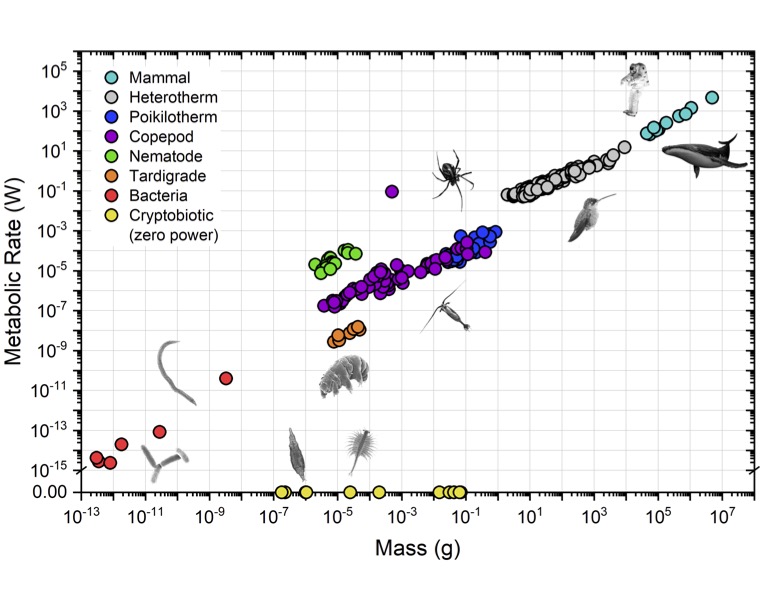}
\caption{\bf{Metabolic rate (MR) and mass for various groups of living organisms.} \normalfont{Despite the vast diversity of species, we observe a near universal energy requirement per unit mass of tissue. This generalization excludes species capable of cryptobiosis (such as tardigrades, brine shrimp, and \textit{Chironomidae}), which exhibit virtually no metabolic activity while in a state of suspended animation, making them better suited for interstellar flight \citep{Geiser2004MetabolicTorpor,Anderson1970MetabolicSpiders.,Anderson1982RespiratorySpiders,Greenstone1980ForagingSpiders,Lighton1995MassStrategists,Clarke1999ScalingFish,Makarieva2005EnergeticsWhales,Zeuthen1947BodyMicrofauna,Heitmann2001MetabolicTemperature,Klekowski1989OxygenSpitsbergen,Ikeda1982OxygenComposition,CLEGG1964TheSalina,GLASHEEN1989MetabolicWater,Crowe1977AnhydrobiosisActivity,Loomis1996MetabolismStates,Okuda2006Trehalose:Vanderplanki,Ricci2008VolumeAnhydrobiosis,Tong2000EffectEdwardsii,Tomlinson2018MeasuringRespirometry,Jonsson2010TrehaloseTardigrades,Anderson1978ResponsesDeprivation,VanAardt2010TheNematode,Wieser1961EcologicalMassachusetts}}.}
\label{fig:metabolic-rate}
\end{figure*}

\section{Species Selection Criteria}

Research success hinges upon practicality and survivability, lending metrics as a basis of comparison between species: low metabolic rate per unit mass, cryptobiotic capability, and high radiation tolerance. 

\subsection{Low metabolic rate}

Launch economics in the near future limit viable organisms to the low-mass regime to ensure achievement of relativistic flight. Due to these engineering constraints, the amount of sustenance we can bring on board is currently severely limited. To survive extended periods of time on a spacecraft where nutrients are scarce, a species will fare better if it exhibits a low metabolic rate as compared to other organisms of similar size. A plot of sub-micron to macroscopic species energy usage per gram of body weight is shown in  Figure~\ref{fig:metabolic-rate}. Species at or near zero energy usage are able to undergo cryptobiosis---a state marked by severely reduced metabolic activity in response to adverse environmental conditions. In order to keep biological systems alive outside of cryptobiosis (e.g., once a spacecraft reaches a target solar system), we require an appropriate environment: pressure, temperature, radiation shielding, nutrients, and waste flow must all be tightly controlled. Current engineering constraints limit the biology available for the interstellar journey, favoring low-maintenance organisms that require less nutrients -- and less payload mass -- to sustain them during the long-duration space travel.

\subsection{Cryptobiotic capability}

To transport life across vast distances in space, an ideal species would undergo cryptobiosis at the beginning of the
mission and would be brought out of hibernation when the destination is reached. Unlike the hibernation of bears or other
animals, which is a lowered metabolic state with energy being consumed from internal storage, organisms in cryptobiosis
maintain extended periods of undetectable metabolic rate. Organisms enter cryptobiosis in response
to extremely harsh environmental conditions such as a lack of water (anhydrobiosis), freezing temperatures (cryobiosis),
fluctuating osmotic conditions (osmobiosis), or a lack of oxygen
(anoxybiosis)~\citep{Withers2018Dormancy,Clegg2002CryptobiosisOrganization}. It is of philosophical
interest to debate whether organisms with no detectable metabolism are ''alive''. To this point, dormant seeds are still
considered alive, though they have largely halted metabolisms until water is imbibed by hydrophilic proteins.
We expect species with cryptobiotic capability to have a higher survival rate during the interstellar travel, as lowered metabolism will confer added protection against the space environment.

Human cryptobiosis has not been achieved but is currently being explored, with avenues such as cryonics and vitrification. Human biology, in addition to many other unfavorable characteristics for long-duration flight, is not well-suited to achieving true cryptobiosis. Of the life forms that are capable of cryptobiosis, only a few are known to be tolerant of spaceflight. These include tardigrades, \textit{C. elegans}, brine shrimp, insects such as \textit{Chironomidae}, most seeds, and microorganisms such as brewer's yeast.

\subsection{High radiation tolerance}

\begin{figure*}[h]
\centering
\includegraphics[width=0.9\textwidth]{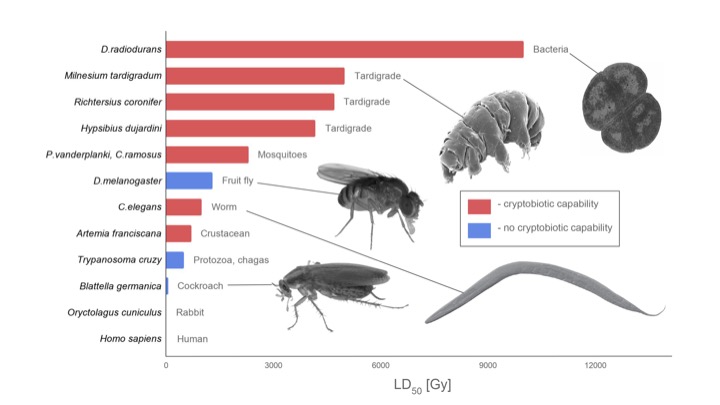}
\caption{\small{In the case of gamma radiation, most organisms exhibit LD$_{50}$ (median lethal dose) values on the order of $10^1$ -- $10^2$ Gy; however, certain nematode, fungi, rotifer, tardigrade, bacterial, and archaeal species demonstrate much higher tolerances ($\sim10^3$ -- $10^4$ Gy) \citep{Jonsson2018EnvironmentalTolerance,Gladyshev2008,Horikawa2006,Jolivet2003,Watanabe2006,Grynberg2012TrypanosomaRadiation,Dvorak2009PossibilitiesArtemia,Jeong2015EffectsPathogens}. Even across species, large differences in LD$_{50}$ values exist, as can be seen between the three represented tardigrade species \textit{M. tardigradum}, \textit{R. coronifer}, and \textit{H. dujardini}.}}
\label{fig:radiation_tolerance}
\end{figure*}

Without the protection of Earth's atmosphere and magnetic field, organisms are subject to much higher fluxes of SCR and
GCR~\citep{Kennedy2014}. The sun produces mostly visible, infrared, and ultraviolet radiation, but higher energetic solar
particle events (SPE) composed of gamma rays, X-rays, protons and electrons do occur. Physical shielding in a tiny
spacecraft can reduce the damage done by ultraviolet radiation, but the other types of radiation produced during SPE can
pass through most lightweight materials, causing radiation sickness and cancer. The vast differences between species'
radiotolerance levels, as shown by  Figure~\ref{fig:radiation_tolerance}, have led to comparative studies to determine
causality. Harrison and Anderson observed four physiological trends in species with high radiotolerance: lower nuclear
material content in cells; the abilities to repopulate cells; regenerate tissue and organs; and the ability to repair
DNA~\citep{Harrison1996}.
The last of these trends may be correlated with desiccation tolerance~\citep{Musilova2015IsolationResistance},
but this is disputed~\citep{BebloVranesevic2018}. Epigenetic factors may also play a role in DNA damage
response~\citep{Montenegro2016TargetingCancer}.
Naturally, the higher the radiation tolerance of an organism, the better it will fare in the space environment. Selecting for species that are better suited to cope with radiation increases the likelihood of success of any on-board experiments. 

High-energy neutrons produced by the interaction of the ISM at the forward edge as well as galactic cosmic radiation with
the spacecraft are also important to consider for interstellar travel. This is discussed in detail in Cohen,
2020~\citep{Cohen2020}. Neutrons can penetrate large distances and damage living organisms~\citep{Kuhne2009}. When these
types of neutrons interact with matter, they can also produce other charged particles and gamma rays. They can cause
clustered and complex DNA lesions that are more difficult to repair than those caused by X-rays, gamma rays, or alpha
particles~\citep{Cucinotta2001}. Numerous studies have reported the induction of mutations, chromosomic aberrations,
reduced survival, cell apoptosis, carcinogenesis, and other effects in multiple different life forms exposed to energy
ranges from 1~MeV to 1000
MeV~\citep{Moreno-Villanueva2017,Kuhne2009,Cucinotta2001,2010IARCHumans,Gersey2007,Nelson1989,IARC2009ACarcinogens}. 

Selecting a species for high radiation tolerance also necessarily includes a discussion of mutation rate. Mutation rate is
defined by the number of mutations that occur per genome replication, and is largely determined by the fidelity of
polymerase activity. There are different types of mutations and correction mechanisms, such as proofreading or mismatch
repair to fix errors, that cause mutation rates to vary extensively across taxa and even across the genome
itself~\citep{Hodgkinson2011VariationGenomes}. Some organisms, such as \textit{Paramecium tetraurelia}, have incredibly
low mutation rates~\citep{Sung2012ExtraordinaryTetraurelia}, whereas viruses like the DNA virus Hepatitis
B~\citep{Caligiuri2016OverviewInfection} have very high mutation rates. As the organisms on board will mostly be in a
cryptobiotic state, the intrinsic rate for each individual species is not of major concern. The concern is the organism's
ability to accumulate mutations from exogenous factors, particularly radiation. The organisms on board should not
accumulate so many mutations during travel as to no longer be viable once the time comes for them to be revived from
cryptobiosis. Even with proper shielding, there will be increased exposure to radiation that may increase the number of
mutations past an unacceptable threshold. In the event this occurs, a species with an already low mutation rate in
addition to DNA repairing capabilities would fare better. An organism like the tardigrade \textit{Ramazzottius
varieornatus} has demonstrated the ability to repair UVC-induced thymine dimers after revival from desiccation and
exposure to fluorescent light~\citep{Horikawa2013AnalysisRadiation}. As such, it is a desirable type of DNA repair and
radiation tolerance mechanism for on-board study. An additional consideration for species selection is the presence of
redundancies in DNA.

\subsection{Selected Species}

\subsubsection{Tardigrades}

Tardigrades, colloquially called ``water bears'' or ``moss piglets'', are microscopic aquatic organisms known for their
robustness. They are found in a variety of habitats ranging from freshwater environments to the ocean and across all seven
continents~\citep{Tsujimoto2016}.
Although not true extremophiles due to their ability to endure but not thrive in harsh conditions, tardigrades have been
shown to withstand droughts, freezing temperatures, high levels of radiation, and high-pressure
environments~\citep{Hengherr2009,Ono2008,Rebecchi2007,Mbjerg2011SurvivalTardigrades}. Tardigrades can enter an
anhydrobiotic state by constricting into a ball known as a tun~\citep{Gagyi-Palffy2011}. The tardigrades' tolerance to
severe environmental factors is contingent on their ability to undergo cryptobiosis. In addition to added protection, the
drastically lowered metabolic needs of the organism while in a cryptobiotic state, suggested to be less than 0.01\% of the
active metabolism~\citep{Piper2007}, make tardigrades ideal for a journey wherein few resources can be available.

Dehydrated and hydrated tardigrades have already been exposed to space conditions in low earth orbit in various
projects~\citep{Jonsson2008,Kennedy2011,Jonsson2016,Persson2011,Rebecchi2009,Rebecchi2011}, and have shown high survival
rates in open space while shielded from UV radiation~\citep{Jonsson2008,Jonsson2016,Rebecchi2011}. Tardigrades are highly
tolerant to $^{56}$Fe ions, a specific type of HZE (high atomic number and energy) particle, and can survive doses as
high as 1000~Gy~\citep{Jonsson2017}. In other organisms, including humans, space radiation and microgravity can initiate
oxidative stress that upsets the homeostasis of the reactive oxidative species (ROS) systems~\citep{Goodwin2018}. In
contrast, the ROS systems of desiccated tardigrades did not show significant differences between samples in space and
controls on earth~\citep{Rizzo2015}. Although desiccation can initiate oxidative stress, it is well-known that tardigrades
can produce the osmolyte trehalose to protect against dehydration~\citep{Hengherr2008TrehaloseDehydration}. Additionally,
tardigrades have intrinsically disordered proteins that generate a glass-like structure during desiccation to help prevent
cellular damage~\citep{Franca2007,Boothby2017}. The tardigrades' proven capacity for maintaining normal system functioning
in extreme conditions may increase its probability of survival for longer periods in space.

The induction or downregulation of certain DNA repair functions, cell cycles, and anti-oxidative stress genes has been
shown to be related to tardigrade response to radiation or
desiccation~\citep{Beltran-Pardo2013,Jonsson2007,Wang2014,Forster2012,Hashimoto2017}. The mechanism responsible for the
increased radiotolerance in tardigrades has been studied in great detail~\citep{Chavez2019}, identifying the role of a
unique nucleosome-binding protein, termed Dsup (damage suppressor). Dsup and its orthologs bind to nucleosomes and shield
chromosomal DNA from radical-mediated cleavage, caused by, but not limited to, radiation from UV rays, X-rays, gamma rays,
and reactive oxidative species. This mechanism of protection has been observed on cultured human
cells~\citep{Hashimoto2016} and more recently in transgenic plants~\citep{Kirke2019}, with transgenic animal models
following promptly. Additional studies are necessary to explore the applications of Dsup to further augment tolerance of
stress-sensitive cells from both plant and animal origins.

\subsubsection{C. elegans}

\textit{C. elegans} is a widely understood multicellular animal whose development and physiology is described at
single-cell resolution. \textit{C. elegans} is optically clear at all stages of its life, allowing the imaging of gene
expression, protein localization, subcellular structures, and cellular physiology in the whole animal at single-cell
resolution by using green fluorescent protein and other spectral variants of fluorescent protein
reporters~\citep{Chalfie1994GreenExpression}. 
The worms are small in size; a teaspoon (5 mL) holds approximately 100 million juvenile worms. They have a very rapid life
cycle, with the laboratory strain of \textit{C. elegans} living about 14 days~\citep{Corsi2006AElegans}.

Like tardigrades, \textit{C. elegans} can be placed in suspended animation by either anhydrobiosis or
cryonics~\citep{Erkut2013MolecularDesiccation}. Indeed, juvenile worms that have been frozen for many decades recover
immediately and proceed to develop to fertile adults that produce hundreds of offspring within two days. Under unfavorable
growth conditions such as starvation or crowding, young larvae enter an alternative developmental stage (the ``dauer''
state) and can survive for several months without
feeding~\citep{Fielenbach2008C.Plasticity,Padilla2012SuspendedQuiescence}. The ability of these organisms to undergo
cryptobiosis -- in addition to the subsequent lower metabolic needs arising from this state -- make them ideal spaceflight
candidates.

Generating transgenic animals in \textit{C. elegans} is relatively straightforward. This has allowed the development of
several transgenic reporters of neural, muscle, and other cellular activity~\citep{KaymakEfficientMosSCI}. There are
trials being conducted to create a transgenic line of \textit{C. elegans} expressing Dsup, to increase resistance to
ionizing radiation, thereby reducing the overall necessary shielding on board the spacecraft. Biological radiotolerance is
the limiting factor in regards to the minimum necessary shielding required aboard the craft. By increasing radiotolerance
genetically, we stand to reduce payload mass considerably.

\subsubsection{Single-celled organisms}

In addition to studying the effects of microgravity and interstellar space radiation on multicellular organisms, analysis
of single-celled organisms is not only possible but presents an opportunity to study evolution in interstellar space. For
example, in akinete (dormant) stages, \textit{Nostoc spp.} have demonstrated desiccation tolerance~\citep{Katoh2003},
resistance to ionizing radiation~\citep{Potts1999}, and survival in vacuum for over a year~\citep{Arai2008}. Previous
experiments on the ISS such as BIOMEX have further validated the survivability of different strains (CCCryo 231-06, UTEX
EE21, and CCMEE 391) of the cyanobacterium in low-Earth orbit (LEO)~\citep{Baque2016}. As terrestrial cyanobacterium,
\textit{Nostoc} is capable of nitrogen and CO$_2$ fixation, as well as oxygen production. 

Studies of \textit{Deinococcus radiodurans}, a bacteria species with one of the highest known radiation tolerances,
suggest that the same mechanisms responsible for repairing DNA after desiccation are also responsible for repairing DNA
after ionizing radiation damage~\citep{Watanabe2006}. In \textit{D.radiodurans}, multiple genome copies, a high degree of
redundancy in DNA repair enzymes, antioxidant defenses, structural toroids, and protein protection with manganese-rich
peptides all play an important role in protecting and repairing radiation damage~\citep{Slade2011,Daly2010,Englander2004}.
Such mechanisms appear unique to this bacteria species, as other organisms demonstrate different tolerance mechanisms.

Another organism, the melanized fungus, has been shown to be highly tolerant of radiation as specimens have been isolated
from irradiated areas~\citep{Zhdanova2004}. The authors report that melanized fungi growth is also stimulated by the
presence of low-dose radiation, with the electron transfer properties of their melanin increasing in radiation's
presence~\citep{Dadachova2007,Robertson2012}. Further studies are necessary to confirm these findings, but the use of
radiation to store metabolic energy presents another avenue for experimental research in the interstellar environment.

In any microbial study persisting over the course of several years,  evolution of the microorganisms and the ability of the bio-containment system to remain axenic will be important. While this may hinder the study of individual species in isolation, an opportunity arises to study the time evolution of community structure and function of microbial communities in the space environment. Such research will aid the development of bioregenerative life support systems (BLSS) that employ microorganisms for O$_2$ production. 

\subsubsection{Other notable organisms and cell types}

Rotifers (particularly bdelloid rotifers) have also been shown to be adept at anhydrobiosis,~\citep{Ricci1987} ionizing
radiation,~\citep{Gladyshev2008} and robustness to low temperatures~\citep{Shain2016}. Indeed, a core sample of
24,000-year old Arctic permafrost revealed rotifers that survived cryptobiosis and were able to reproduce (via
parthenogenesis)~\citep{Shmakova2021}.

Mammalian cells are not as robust to the space environment as the previously mentioned organisms. However, understanding
the extrasolar space environment's effects on mammalian cell development and division is one of the first steps to sending
additional life forms out of the solar system. Microgravity and ionizing radiation has shown to have degenerative effects
on a range of human stem cells~\citep{Blaber2014,Cerutti1974},
but mutation research suggests that mammalian cells may be able to produce an adaptive response when exposed to low levels
of radiation that then strengthens their resistance to subsequent higher levels~\citep{Vijayalaxmi2014}. More research is
needed before we can make confident statements concerning the effect of space conditions on mammalian cells. Human
embryonic kidney (HEK) cells, Chinese hamster ovary (CHO) cells, and fibroblasts are all favorable subjects for these sort
of tests~\citep{Niwa2006}, several of which are ongoing in LEO.

\section{Experimental Opportunities}

\subsection{Range of biology to be tested}

Selecting for species that are physiologically better fit for interstellar travel opens up new avenues for space research.
In testing the metabolism, development, and replication of species, like \textit{C. elegans}, we can see how biological
systems are generally affected by space conditions. \textit{C. elegans} and tardigrades are inherently more suited to
space flight as opposed to humans due to factors like the extensive DNA protection mechanisms some tardigrades have for
radiation exposure or the dauer larva state of arrested development \textit{C. elegans} enter when faced with unfavorable
growth conditions. However, as there is overlap between our species, like the human orthologs for 80\% of \textit{C.
elegans} proteins~\citep{Lai2000}, we can begin to make some predictions on the potential for human life in interstellar
space. 

Although the current size of the interstellar spacecraft restricts the species available for study, the use of \textit{C. elegans} and tardigrades will be able to return statistically significant results. Even if radiation-induced population losses occur, the remaining worm and tardigrade count available onboard is still expected to be enough to conclude on findings with significance. In addition, \textit{C. elegans} can easily be cultured on a diet of \textit{E. coli} bacteria or completely defined synthetic growth medium while tardigrades can feed on \textit{C. elegans}. This provides us with the opportunity to study the effects of interstellar travel on reproduction, growth, and development of an organism in addition to the effects on a food chain in statistically significant numbers.

Given the desire for low power and mass on board the spacecraft, new designs for low mass fluorescent imaging may be needed. We can also harness the ease of transgenesis and optical properties of \textit{C. elegans} to develop transgenic worms with bioluminescent reporters for muscle and neural activity. In addition, we will develop reporters to monitor DNA repair response to exposure to cosmic radiation and study its mutagenic effect and evolution during interstellar space travel. Thus \textit{C. elegans} will provide an unprecedented window into the effects of hypergravity, microgravity, and cosmic radiation associated with interstellar travel on physiology and anatomy of a whole organism at single-cell resolution.

An interesting avenue of research that can be explored is memory retention, specifically in \textit{C. elegans}; the worms
can be trained to associate odors with food and exhibit classic Pavlovian associative learning
behaviors~\citep{Rothman2019}. It has been shown long-term associative memory lasts up to 40~h with noticeable declines
beginning after 16~h~\citep{Kauffman2011}. If proven to be capable of memory retention after extended periods of
cryptobiosis, \textit{C. elegans} can be used in a variety of experiments with the limiting variable as time. Testing is
currently being done to determine the tardigrades' capacity to be classically conditioned.

\subsection{Experimental design}

To house specimens on board, we are developing the design of a simple microfluidics chamber that provides  conditions
necessary for reviving and sustaining life (microorganisms of order 200 $\mu$m in length). These dimensions
are large enough to perform remote lab-on-a-chip experiments, but also small enough to be viable for low mass spacecraft.
Polydimethylsiloxane (PDMS) is the typical material in microfluidic applications due to its biocompatibility,
transparency, and ease of fabrication~\citep{Roy2016}, but it is not ideal for the space environment. PDMS has several
disadvantages, including high O$_2$ permeability, strong adsorption of biomolecules, relatively high glass
transition temperature, and poor mechanical compliance. New thermoplastic elastomers are compatible with microfabrication
and biology, mechanically stable, less permeable to O$_2$ than PDMS, and optically transparent (UV to IR
range)~\citep{Lachaux2017ThermoplasticDevices}.
They can be manufactured with diverse glass transition temperatures and either monolithically prepared with an imprinter
or integrated with other candidates, such as glass. Polymerase chain reaction (PCR) on a chip is another area that will
evolve naturally over the next decade and is one of many biological techniques that could be incorporated in designs. It
is also possible that enzymes could be stabilized in osmolytes to perform onboard biochemical reactions. For the study of
life in the interstellar environment, it is necessary to include experimental controls (in LEO and on the ground) and the
use of diverse genotypes and species to characterize a wide response space~\citep{Vandenbrink2016}.
Initial research on miniaturizing the necessary research hardware and supporting electrical
components is detailed in Brashears, 2016~\citep{Brashears2016}.

\subsection{Abiogenesis experiments}

A hermetically sealed environment aboard the
spacecraft would enable unprecedented abiogenesis experiments with the ability to conduct further research on prebiotic
chemical reactions. The primary engineering constraint will be the source of energy, as it must be able to emulate an
Earth-like environment for the entire duration of the experiment. Biologically, the challenge rests with forming nucleic
acids and other key biological molecules out of the various complex organic molecules expected in order for life as we
know it, to arise---a feat that has only been recreated under tightly controlled
conditions~\citep{Gibson2010CreationGenome}. The high radiation environment of interstellar space provides an interesting
opportunity to study the biochemical origins of life in spacecraft with low radiation shielding versus those equipped with
protective measures to limit the effect of galactic cosmic radiation (GCR) on the prebiotic chemicals, possibly shedding
light on the flexibility in the conditions needed for life to arise.

\subsection{An interstellar seed and genetic ark}

Much like the seed banks on Earth such as the Svalbard seed vault that acts as a ``backup'' in case of a large-scale disaster on Earth or in the solar system, we can conceive of variants of the missions we propose as the ultimate backup of life on Earth. Seeds have been shown to be able to survive for extremely long times, with 32,000 year old seeds being viable. The idea of storing and propagating the ``seeds of life'' is a natural extension of our program. In this case, ``seeds'' can be physical seeds of all kinds from plant to human as well as the genetic precursors. Slower-moving payloads can potentially be recovered by future technologies or extraterrestrial intelligence.

\subsection{Digital ark --- backup of Earth}

In addition to the physical propagation of life, we can also send out digital backups of the ``blueprints of life'', a
sort of ``how-to'' guide to replicating the life and knowledge of Earth. The increasing density of data storage allows for
current storage density of more than a petabyte per gram and with new techniques, such as DNA encoding of information,
much larger amounts of storage can be envisioned. As an indication of viability, we note the US Library of Congress with
some 20 million books only requires about 20 TB to store. A small picture and letter from every person on Earth, as in the
``Voices of Humanity'' project, would only require about 100 TB to store, easily fitting on the smallest of our spacecraft
designs. Protecting these legacy data sets from radiation damage is key and is discussed in Lubin 2020 and Cohen et al.
2020~\citep{Lubin2020,Cohen2020}.

\section{Planetary Protection for Extrasolar Biology Missions}

\subsection{Relevance of planetary protection regulations and measures taken}

Sending Earth-based life to interstellar space using a directed-energy powered craft requires addressing risks of possible
contamination of extrasolar planets. As planetary protection guidelines under Article IX of the Outer Space
Treaty~\citep{1967TreatyBodies} are until now only reified for solar system bodies in the Committee on Space Research
(COSPAR) regulations (2011), the first part of the mission -- moving through our own solar system -- is of primary concern.
COSPAR's international planetary protection guidelines regulate missions' profiles and their potential to contaminate
life-supporting areas of solar system bodies with Earth organisms (forward contamination)~\citep{Sherwood2019}, and of
possibly contaminating Earth through sample return missions (backward contamination). The proposed interstellar mission
profile will only touch the aspects of forward contamination, as it has no plans of collecting and returning samples. 

A comparable example of a mission with biological payload is the LIFE Experiment aboard the failed Phobos Grunt Mission.
Although all planetary protection regulations were followed~\citep{Khamidullina2012RealizationMission}, considerable
criticism from different sources followed~\citep{DiGregorio2009InternationalMars}.
Criticism of this mission profile concerning the potential for forward contamination by the failed
mission~\citep{DiGregorio2010DontMars} is thus not applicable for the reasons already outlined above. A suggested
fail-safe mechanism in the form of a self-destruct mechanism is already integrated into this mission by the speed of the
craft.

Independent of the actual scope of the regulation for biological forward contamination, the suggested mission profile will live up to the standards. An object with a mass of less than ten grams accelerating with potentially hundreds of GW of power, will, even if it were aimed at a planetary protection target (for example Mars), enter its atmosphere or impact the solar system body with enough kinetic energy to cause total sterilization of the biological samples on board. The velocity of the craft would thus serve as an in-built mechanism for sterilization. The mission profile does not include deceleration, so this mechanism is valid for the entirety of the mission. Apart from sterilization, the spacecraft will be aimed at a target outside of the ecliptic and will be accelerated rapidly by directed energy. The probability of an incident relevant to planetary protection regulations (e.g.~inadvertently hitting a planet on the way out of the solar system) is thus significantly lower than the regulatory probability of $10^{-6}$. In summary, the requirements for demonstrating probabilities of 99\% to avoid impact for 20~years, and 95\% to avoid impact for 50~years, will be fulfilled by the velocity of the craft and the target lying outside of the solar system.

The intended final velocity of the spacecraft depends on the spacecraft's mass and the DE system's power and size. One goal of Starlight is to achieve speeds greater than 25\% of the speed of light. Upon impact, the kinetic energy (${\sim}1$ kiloton/g at 0.25$c$) will sterilize the entire spacecraft and adhere to COSPAR planetary protection regulations outside of the solar system as well, even if those regulations are not applicable at this time. Should planetary protection regulations be extended to extrasolar missions, a mission type such as the one proposed in this paper will automatically adhere to any regulations regarding probabilities of extrasolar forward contamination.

\subsection{Extrasolar protection looking forward}

The discussion of extrasolar planetary protection rules by Charles Cockell concerning sterilization of craft by long
exposure to the ISM on the example of the \textit{Voyager} craft makes an interesting point for DE
missions~\citep{Cockell2008}, Cockell argues that long-duration missions with lifespans larger than $10^3$ years will
automatically be sterilized through interstellar radiation. If the proposed profile is a flyby mission, the mission
lifespan will automatically stretch to larger than $10^3$ years, despite the target being reached in decades, as the
craft will enter the target solar system for a period of time before flying on. If the spacecraft entered the atmosphere
of an exoplanet, the craft's kinetic energy will ensure a similar or greater level of sterilization than the kind
suggested by Cockell~\citep{Cockell2008}. However, the released kinetic energy poses other ethical problems beyond
contamination by Earth life. The energy released will be in the range of $10^{12}$ J or about a kiloton TNT equivalent,
similar to measured impact energies from large meteorites and small asteroids. For a planet with an atmosphere similar to
Earth, this would be unnoticeable except to extremely sensitive instruments. Schwartz argues for the ethical values of
uninhabited or uninhabitable places in the solar system and beyond~\citep{Schwartz2019}, drawing on Cockell and Horneck's
arguments for planetary park systems~\citep{Cockell2006,Cockell2004}. The park system would seek to preserve a planet's
pre-existing environment, just as national parks do on Earth. This would pose a challenge for any directed-energy mission
entering a planetary atmosphere at a significant fraction of the speed of light, regardless of its payload. 

The planetary park system demonstrates a need to reevaluate planetary protection considerations and regulations in light
of coming extrasolar missions. While planetary protection -- especially in the form of forward contamination regulations
for solar system bodies -- aims to protect scientific research from unintended contamination, this rationale no longer
works for extrasolar missions. Outside the solar system, distances are so vast and the mission's timescale is so large
that the protection of science cannot be the only rationale \citep{Gros2019}. The value of sending
Earth life to other solar systems needs to be discussed to reform planetary protection regulations. The ethical value of
paving the way for future human missions to other solar systems through biological precursor missions like this one cannot
be addressed within current regulations. The lead author subscribes to Schwartz's train of thought
for interstellar planetary protection in seeking to preserve exoplanets beyond the life bias, including preventing
contamination during ``first visits''~\citep{Schwartz2019}. However, once initial exploratory phases have been completed
and best efforts at gathering knowledge of exoplanets has been performed, there may be certain circumstances in which
intentionally ``seeding life'' would not be unethical.

The idea of ``seeding life'', whether accidentally or intentionally on an
exoplanet~\citep{Gros2016}, is intriguing. Re-entry, assuming there is an atmosphere or direct impact to the surface if
there is no atmosphere, requires the dissipation of an extremely large amount of energy per mass (about 60~MeV per nucleon
of about 1 kiloton TNT/gram at 0.33$c$). Coupling this amount of energy directly to the spacecraft with no other
dissipation modes will result in complete vaporization of the spacecraft. Various techniques for slowing the spacecraft
have been discussed \citep{Lubin2020,Gros2017}, but none of them appear to be practical at the
extreme speeds we are proposing. There are some possibilities for speeds less than about 0.05$c$ including using
photon braking from bright stars. For now, reentry remains an open question. 

\section{Conclusion}

We are rapidly approaching the technological capability for interstellar flight on meaningful timescales. As such, we must
consider the benefits of sending life outside of the solar system. Using remote sensing turned inward, we can study
biological systems under microgravity and space radiation conditions unique to interstellar space as well as remote
sensing of the interstellar target. Interstellar probes may yet bring us closer to answering questions long pondered in
the stories of science fiction, such as ``Can humans travel to other star systems?'', but may also allow us to investigate
more pragmatic matters, such as the plausibility of the panspermia hypothesis of the origins of life on Earth. Exploration
via interstellar probes is as such warranted; the United States Congress has even encouraged NASA to develop a conceptual
roadmap for doing so~\citep{2016CommerceCong.}.

Here we have outlined several challenges associated with studying terrestrial life in interstellar space, namely transport, data acquisition, life support, experimental design, and ethical considerations. While this list is by no means exhaustive, we see these as necessary considerations for producing meaningful research results and urge the community to foster more development in these areas. We see the standoff directed energy propulsion technology from Project Starlight as a possible option for interstellar transport of biological systems. Species selection criteria have also been identified, as well as examples of species suitable for experiments. Sending small (micro-scale) cryptobiotic lifeforms as model organisms can pave the way for researching more complex, less robust organisms in interstellar space. The information presented in this paper aims to facilitate the conceptual design and elucidate the experimental space currently available to interstellar space biology missions. Future work in the development of more sophisticated interstellar biosentinel experiments and hardware design, eventually culminating in a deeper understanding of both our own biology and worlds beyond, will be one of humanity's greatest endeavors.

\section*{8. Acknowledgments} 
Funding for this program comes from NASA grants NIAC Phase I DEEP-IN [NNX15AL91G] and NASA NIAC Phase II DEIS [NNX16AL32G] and the NASA California Space Grant Consortium [NNX10AT93H] as well as a generous gift from the Emmett and Gladys W. fund (P.L.). Additional funding comes from the National Institutes of Health [1R01HD082347 and 1R01HD081266] (J.H.R.).

 \bibliographystyle{elsarticle-num} 
 \bibliography{references2-2}





\end{document}